# AI Failures: A Review of Underlying Issues

**Debarag Narayan Banerjee**
Head, Data Science, *Agoda*
debarag.banerjee@gmail.com

**Sasanka Sekhar Chanda**
Professor, Strategic Management, *IIM Indore*
sasanka2012@gmail.com

**Abstract.** Instances of *Artificial Intelligence* (**AI**) systems failing to deliver consistent, satisfactory performance are legion. We investigate why **AI** failures occur. We address only a narrow subset of the broader field of *AI Safety*. We focus on **AI** failures on account of flaws in conceptualization, design and deployment. Other *AI Safety* issues like trade-offs between privacy and security or convenience, bad actors hacking into **AI** systems to create mayhem or bad actors deploying **AI** for purposes harmful to humanity and are out of scope of our discussion. We find that **AI** systems fail on account of omission and commission errors in the design of the **AI** system, as well as upon failure to develop an appropriate interpretation of input information. Moreover, even when there is no significant flaw in the **AI** software, an **AI** system may fail because the hardware is incapable of robust performance across environments. Finally an **AI** system is quite likely to fail in situations where, in effect, it is called upon to deliver moral judgments—a capability **AI** does not possess. We observe certain trade-offs in measures to mitigate a subset of **AI** failures and provide some recommendations.

Artificial intelligence (**AI**) systems seek to employ computing machines to carry out tasks requiring reasoning, knowledge representation, planning, learning, natural language processing, and perception and tasks involving moving and manipulating objects [1]. In this study, we present causes of **AI** failure grouped under stylized headings and provide recommendations to address certain failure situations. It is hoped that the study will influence development of better **AI** systems, as well as encourage pursuit of non-**AI** solutions where fatal **AI** flaws cannot be avoided.

**1.0. Omission errors in the design of an AI system**. This happens when the **AI** system is unable to draw on a pre-configured recourse to handle a situation it encounters under conditions of live operation. One such instance of **AI** failure was observed in Las Vegas [2]. A driverless shuttle bus stopped but failed to sound the horn to alert the driver of a delivery truck that was reversing on to a lane perpendicular to the bus's path. Eventually the truck's wheel clipped the fender of the shuttle.

Another flavor of omission error originates in failure to design the **AI** system to cope appropriately with certain variations in a known scenario. For example, the "liveness" detection feature of iPhone's *Face ID*—which is used to confirm the person *Face ID* is looking at is real and not a mask or someone wearing prosthetics—was defeated with glasses and tape [**3**]. Relatedly, it has been shown [**4**] that in Amazon's *Rekognition* tool, photos of senators of color are more likely to be misidentified as matching with mugshots of persons arrested on suspicion of criminal conduct. It is probable that the 'face template' used in the *Rekognition* tool has a large number of parameters on which *only* non-colored people (e.g. white males) exhibit significant variation. Omission of parameters in which the colored people exhibit variation leads to course-grained judgments, resulting in higher extent of false positives.

**2.0. Commission errors in the design of an AI system**. This happens when an **AI** system takes an inappropriate action on account of flawed design of the command-processing module of the **AI** system. One kind of commission error is manifested when the **AI** system incorrectly construes an environmental event as a command for action. The **AI** system ends up taking an action *not* requested by the end-user / owner. This error happens when a *wake-up* or *take-action* command is not sharply distinguished from environmental events. In Portland, U.S.A., Amazon *Echo* listened to a couple's conversation about hardwood floors and sent the recording to someone in their contact list—without the couple's knowledge [**5**]. A variation of this commission error manifests when the **AI** system takes action when it is not asked to, and such happens when there are nil environmental cues to misconstrue. This is likely to happen when the logic in the **AI** system's internal processing routines generates a *wake-up* or *take-action* command inappropriately. Moreover, on some occasions, the introduction of new features (or 'updates') breaks the robustness of existing features, in that the functioning of a new feature (or 'update') itself—and not environmental noise—inappropriately directs the **AI** system to take action. An instance of an **AI** system taking action without being asked to is observed in the case where Amazon *Echo* turned on music from *Spotify* at full volume, at an hour past midnight, in an empty sixth floor apartment in Germany [**6**].

**3.0. AI failure arising from inappropriate interpretation of input information**. This happens when an **AI** system takes inappropriate action upon failing to interpret input information correctly. Shortly after its release in 2011, Apple's *Siri* agreed to memorize the name of its owner as "an ambulance", simply because the latter had issued a command "Siri, call me an ambulance." [**7**] *Siri* failed to disambiguate between alternate uses of the phrase "call me". In eastern China, a traffic camera using **AI** deemed that a driver was "driving while holding a phone"—an offence meriting a fine—when the corresponding picture clearly shows that the driver was just scratching his face [**8**].

**4.0. AI fails because the hardware is incapable of robust performance across environments**. The hardware of an **AI** system comprises not just the CPU, GPU and servers in the cloud, it also involves sensors including image capture devices like camera, devices to capture sound, temperature, humidity etc., and related wiring to the computational device(s) running **AI** software. The degree of robustness of functioning of the hardware and sensors feeding data to an **AI** system—under a range of environmental conditions pertaining to fluctuation of temperature, humidity, dust and suspended particulate matter, lighting, foreground and background noise, incidence of electric and magnetic waves—impact **AI** functioning even when there are no major flaws in the **AI** software. For instance, in Staten Island, New York, a ten-year-old boy was able to unlock his mother's iPhone X using his own face for *Face ID* verification, under certain lighting conditions in one of the family's bedrooms [**9**]. The Boeing Max crashes (*Lion Air* Flight 610 in October 2018 and *Ethiopian Airlines* Flight 302 in March 2019) had an **AI** system (MCAS) dangerously bending the nose of the airplane downwards regardless of altitude—in a misdirected bid to avoid stall as happens if the plane had its nose raised too far—based on reading from just one (angle-of-attack) sensor [**10**]. Erratic sensor outputs doomed the two fatal 737 Max crashes because it left the pilots fighting the control system and losing the battle, but the problem could have been either the sensor or in the wiring that connects the sensor to the flight computer and thence to the flight software and MCAS. One viewpoint is that if MCAS was not rushed through development, it could have taken input from *other sensors* as well, in order to arrive at a flight solution, such as from the plane's artificial horizon, airspeed indicator, altitude data and other computer solutions on the flight profile.

**5.0. AI fails because it is unable to make moral judgments**. This point addresses situations where providing more data, improved design and/or better hardware is not going to make problems in **AI** functioning go away. When an **AI** system encounters a novel situation for which pre-configured rules cannot be readily found or for which contradicting rules seem to apply, the **AI** system either fails to take action, or chooses randomly from the set of actions feasible, or takes the first action it reads from its code, or repeats a previous action. Alternately, the **AI** system may take an action that addresses the first (or last) sub-set of conditions for which built-in processing routines exist. The response from the **AI** system is likely to be inappropriate in most cases. In contrast, humans are often able to use their moral judgment to determine appropriate actions in unanticipated situations. Morality takes root in humans through (a) conditioning since childhood and over generations, shifting as social mores shift, and (b) by observation and learning from the action of role models in various situations. Human decision-making is nuanced by virtue of richer input and richer repertoire of processes to interpret input and construct a response. A human receives information from many channels—sight, sound, smell, touch, taste, as well as by observing reactions of other humans—and draws from a wealth of prior experience. Humans are able impute meaning to actions and gestures of others, taking into account the context as well as the cultural setting. Human conversation and gestures convey much more than what is embodied when the words are codified into text and symbols. **AI** has access only to the codified texts and symbols. The design of an **AI** system admits to a less complex interlinking relationship among channels providing inputs. One and only one meaning is imputed to a particular signal or pattern of signals, by **AI**. The nature of clarifying questions asked by **AI** (to human end-users) is also rather primitive, compared to the nature of clarifying questions a human can ask another human, *in order understand a situation better*. **AI** cannot succeed in situations where only the human can draw from behavior, speech, actions and non-verbal cues arising in interactions with other humans, as well as where only the human can draw from prior socialization experiences and other contextual information and interpretations thereof. The following incident featuring *Sophia* the humanoid robot is instructive:

In March of 2016, Sophia's creator, David Hanson of Hanson Robotics, asked Sophia during a live demonstration at the SXSW festival, "Do you want to destroy humans? ...Please say 'no.'" With a blank expression, Sophia responded, "OK. I will destroy humans." [**11**]

Sophia misinterpreted a request to provide her opinion about something as a request to agree to carry out the activity in question. Sophia interpreted the "Please" after Hanson's question as a request to agree to do as "asked" in the previous sentence (i.e. agree to destroy humans), and failed to comprehend that the full sentence—"Please say 'no'"—requested just the opposite. We note that Sophia did not directly answer whether she *wants* to destroy humans. Sophia's response ("OK. I will …") suggests she merely agreed to carry out a command given to her. In general **AI** shall find it difficult to resolve the clash between two competing values (a) to always cooperate with the human owner and (b) to not harm humans. In contrast, humans are able to choose among two alternatives by moral reasoning, on most occasions.

**Debrief and Recommendations.** Given the state of technology today, we feel that **AI** should not be deployed in frontline roles that require making moral judgements that drive rewards and punishment for humans (# **5.0.** above). To address # **4.0.**, triangulation from multiple sources should be insisted upon, to bolster reliability of **AI** hardware. Moreover, there is need for an ongoing discussion in the society regarding the circumstances under which mass deployment of cheap low-capacity hardware (for instance low-resolution cameras)—that yield questionable evidence—is preferable over using human intelligence and situation-awareness. Further, in situations requiring disambiguation of human interactions—speech, symbols and behavior (# **3.0.** above)—we feel that trained human agents must be deployed complementing **AI**.

In order to address omission errors where certain a pre-configured recourse to handle a situation is missing (# **1.0.** above), we need to recognize an underlying trade-off. Take the case of driverless vehicles: Rides by driverless-vehicles will tend to be jerkier when the **AI** system pays attention to a greater number of "threats" and slams brakes to wait till a threat goes away; a smoother ride is possible only when a majority of threats are ignored by the **AI** system, but it may lead to serious accidents, as happened when an **AI**-driven *Uber* vehicle hit and killed a pedestrian [**12**]. The only reasonable solution appears to be to have driverless vehicles ply under the watch of engineers/ emergency drivers *sitting inside the car* for as long as it takes to develop confidence regarding safety. For other **AI** applications, the trade-off is between **AI** requesting inputs from the human owner annoyingly frequently, vs. the **AI** system failing to ask in critical cases.

There is also a trade-off underlying the situation where variations of some known scenarios are missing (# **1.0.** above): **AI** designers devote more time to think up variations, lengthening development time (while not escaping groupthink) vs. putting the **AI** solution "out there" to generate new variations, to be taken care of, subsequently. Here again, we feel that an **AI** system should not function un-chaperoned in applications that potentially involve unconscionable levels of risk to human safety, a case to point being the instance a worker was killed by a robot in a Volkswagen plant in Germany [**13**].

In order to rectify the situation where an **AI** system responds to environmental cues instead of responding *only* to owner commands (# **2.0.** above), further research on biometrics is needed to make the owner's command unique, but not so unique as to lock the owner out from getting response when his/her voice changes due to age, illness etc. Alternately, separating the owner and the command generation mechanism may help. For example, the owner may initiate a sound of a certain frequency by a *device*, to wake up an **AI**-assistant. However, this option could be problematic if the device to make the wake-up sound falls in wrong hands. Further, much needs to be done to prevent **AI** from acting when owner-generated and environment-generated cues are absent. A centralized control of the **AI** design and development process may help. But it will slow the pace of new releases, since every new release must be tested for potentially breaking all prior functionality. This gets complicated when **AI** systems interact with other third-party systems, for example, *Echo* connecting to *Spotify*. Unless rigid input-output protocols are maintained, a new release of functionality of *Echo* may not be protected from every potential idiosyncratic behavior of *Spotify* and vice-versa.

In sum, we observe that just as all contracts are incomplete[1]—or, alternately, just as no sample can represent a population fully—**AI** developers may fail to foresee all manner of contingencies that can arise in *live* operation. Alternately, a commercial imperative to get an **AI** solution out of the door as quickly as possible can also lead to problems arising from (a) insufficient number of scenarios being considered, and (b) certain variations of known scenarios getting missed out. We observe another compounding factor compromising **AI** development: thinking in silo-s. For example, in the case of the MCAS system indicted in the Boeing Max accidents, we are otherwise at a loss to explain that a reliability engineer did not object to tying the MCAS to just one sensor, and thereby the specification to consider having MCAS act on the basis of readings from multiple sensors never reached the **AI** development team. An alternate explanation—that for MCAS, single sensor reading was preferred over multiple sensor readings *merely* to reduce cost and/or delivery time—appears far-fetched given Boeing's rich engineering heritage. Moreover, the design outcome that the MCAS system can wrest control of the plane from the human pilot and point the nose of the plane downwards *even at low altitudes* is equally inexplicable, unless we impute presence of unconscionable extent of silo-d thinking.

Last but not the least, we see tremendous potential for developing **Localized AI solutions** —**AI** solutions that don't require tethering to servers in the remote cloud—in the same way the desktop revolution provided an alternative to the dumb-Terminal-smart-Server paradigm dominant in an earlier era. At present a vast majority of the **AI** systems are designed to interact with the cloud. The primary purpose is to gather as much user and usage data as possible for a range of purposes, current and future, some noble, some not so. When we transcend the mind-block of endeavoring to make every **AI** application tied to the cloud and see what can be done better, locally, a range of new **AI** applications come to the fore. For example, deep learning and allied technologies may be deployed to study how cells regenerate in humans, and tasked with re-growing a damaged organ. The alternative in use today—organ transplants—has low success rates and ties patients to medications for the rest of their lives. Likewise, **AI-**based learning may be deployed to extract a significant proportion of the suspended particulate matter (SPM) from mines and from the flue gas of thermal power-plants and oil rigs—perhaps converting the SPM into useful fertilizer or building material—lowering the incidence of respiratory diseases in the neighborhood. Further, the productive life of solar panels may be enhanced by deploying **AI**-based learning to clean cells in real-time, arresting the depletion of the height of the potential barrier, delaying the aging process of cells.

---

[1] We learn from the research of Grossman and Hart [**14**] that no contract between two parties—say between the **AI** developer and the **AI** end-user—can specify what is to be done in *every* possible contingency.